\documentclass[showkeys,showpacs]{revtex4-1}

\draft 
\usepackage{color}
\usepackage{amsmath}
\usepackage{amssymb}
\usepackage{graphics}
\usepackage{graphicx}
\usepackage{subfigure}
\usepackage{natbib}

\begin{document}


\title{Pauli Oscillator In Noncommutative Space} 



\author{Mebarek Heddar}
\email[mebarek.heddar@univ-biskra.dz]
\thanks{}
\affiliation{Laboratory of Photonic Physics and Nano-Materials (LPPNNM), Depertment of Matter Sciences, University of Biskra, Algeria.}

\author{Mokhtar Falek}
\email[mokhtar.Falek@univ-biskra.dz]
\thanks{}
\affiliation{Laboratory of Photonic Physics and Nano-Materials (LPPNNM), Depertment of Matter Sciences, University of Biskra, Algeria.}

\author{Mustafa Moumni}
\email[m.moumni@univ-biskra.dz]
 \thanks{}
\affiliation{Laboratory of Photonic Physics and Nano-Materials (LPPNNM), Depertment of Matter Sciences, University of Biskra, Algeria,} 
\affiliation{
PRIMALAB, University of Batnal, Algerie.}

\author{Bekir Can L\"utf\"uo\u{g}lu}
\email[Corresponding author:bclutfuoglu@akdeniz.edu.tr]
\thanks{}
\affiliation{Department of Physics,  Faculty of Science, Akdeniz University, 07058 Antalya, Turkey, }
\affiliation{Department of Physics, University of Hradec Kr\'alov\'e, 	Rokitansk\'eho 62, 500 03
Hradec Kr\'alov\'e, Czech Republic.}

\date{\today}

\begin{abstract}
In this study, we investigate the Pauli oscillator in a noncommutative space. In other words, we derive wave function and energy spectrum of a spin half non-relativistic charged particle that is moving under a constant magnetic field with an oscillator potential in noncommutative space. We obtain critical values of the deformation parameter and the magnetic field,  which they counteract the normal and anomalous Zeeman effects. Moreover, we find that the deformation parameter has to be smaller than $2.57\times 10^{-26} m^2$.  Then,  we derive the Helmholtz free energy, internal energy, specific heat and entropy functions of the Pauli oscillator in the non commutative space. With graphical methods, at first, we compare these functions with the ordinary ones, and then,  we demonstrate the effects of magnetic field on these thermodynamic functions in the commutative and noncommutative space, respectively. 
\end{abstract}

\keywords{Pauli oscillator; noncommutative space; thermal properties}

\pacs{03.65.Pm, 03.65.-w, 03.65.Ge}

\maketitle
\date{\today}
\section{Introduction}\label{Int}
The first proposal of the presence of a  noncommutative (NC) spacetime goes back to the middle of the last century. Regarding ultraviolet divergences found in quantum field theory, Snyder proposed NC operator algebra of the quantized spacetime coordinates  \cite{Snyder1947}. In the meantime, renormalization techniques were developed so that Snyder's groundbreaking work began to attract less attention. In the last decade of that previous century, interest has been renewed by the discovery of Seiberg Witten map \cite{Seiberg99}, which allows to associate NC gauge theories  with commutative ones \cite{Kazanawa2019}. In the following decade, many papers are written on the NC field theories \cite{T1NC,T2NC,T3NC,T4NC,T5NC,T6NC,T7NC}, including the possible experimental investigations \cite{Ex1NC,Ex2NC}. For further reading, we refer the following reviews \cite{Rev1,Rev2,Rev3}. It should be noted that studies on string theory have also increased the interest  to the NC spacetime \cite{Connes98, Ardalan99}.  The appearance of NC geometry effect in the very tiny string scales \cite{Witten96}, M-theory \cite{Banks97,Nair01} compactification and quantum Hall effect \cite{Susskind01, Hellerman01} were the main motivation of these studies. Non-commutativity  in quantum field theory can be achieved in either of two different ways: via using Moyal product on the space of ordinary functions, or by defining the field theory on a coordinate operator space that is intrinsically NC \cite{Douglas01}. Evidence of equivalence between the two approaches is discussed in \cite{Alvarez01, Akofor}. 

In the meantime, these intriguing discussions motivated people to investigate the noncommutativity at the level of quantum mechanics \cite{Gamboa012, Belluci01, Chaichian01, Gamboa02, Demet02, Moumni11,HH01,HH02,HH03, Boumali01, Boumali02,Zhong2021}. We observe that different potential energies are taken into account to understand the physical effect of NC space. For example, the isotropic harmonic oscillator potential energy are considered in \cite{Hatzinikitas02, Smailagic021, Smailagic022, Kijanka04,BenGeloun09} to obtain the bound state solutions in two and three dimensions. Dirac oscillator which is mainly considered to describe the quark-confining processes in quantum chromodynamics \cite{Ito, Moshinsky89}, is examined in NC phase space in \cite{Mirza04,Cai10,HB2013,Hou2015,Sargol}. The Schrodinger oscillator which is used to describe the confinement of quarks in mesons and baryons, are examined in NC phase space in \cite{Santos2011}. The Klein-Gordon oscillator solution in NC geometry is given in \cite{Mirza04, Wang2008,Zaim2014}.  Alike, the Duffin-Kemmer-Petiau oscillator is investigated  for spin-0 and spin-1 particles in NC phase space to comprehend the NC effects on the energy spectrum \cite{FalekMerad2008,FalekMerad2009,Guang2009}.
 
Besides of these oscillators, there is another interesting non-relativistic oscillator, namely the Pauli oscillator (PO) \cite{Pauli1927}. Basically, it is used to describe the dynamics of the charged spin-half particle according to the particle's spin interactions with an external electromagnetic field. The bound state energy spectrum levels of a PO are known as the Landau levels. To our best knowledge, there are several works which intend to explain the Landau problem in the NC geometry. In one of them, it is shown that the equation of motion of the harmonic oscillator in a two dimensional NC space produces a similar result in the lowest Landau level, to that of a particle's one in a constant magnetic field in the commutative space \cite{Gamboa011}. After that, Mirza \emph{et al.}  showed this equivalence is valid in the relativistic sector, namely in the Dirac and Klein-Gordon oscillators, although an exact map does not exist \cite{Mirza04}. Recently, Haouam studied the two and three dimensional Pauli equation in NC phase-space \cite{Haouam20a,Haouam20b}.

The aim of this work is to obtain an exact solution to the Po problem in the NC space, and then, to examine the effect of the deformation on the thermodynamic properties of a non-relativistic system with spin. The outline of this paper is as follows: In section \ref{Pauli}, we expose the analytic solution of the Pauli equation of oscillator in the NC space. First, we introduce PO Hamiltonian in commutative space, then we extend it to the NC space. After that we obtain the energy spectrum and  investigate the critical cases where the effect of magnetic field is eliminated by the NC effects. Our results set an upper bound value for the deformation parameter. In section \ref{Thermal}, we derive the thermodynamic properties of the PO in the regime of high temperatures. Moreover, we present discussions on the thermal properties and support them with their demonstrations. Finally, in the last section we give our concluding remarks.

\section{Pauli oscillator}\label{Pauli}
One can base the Pauli Hamiltonian  on the non-relativistic limit of the Dirac Hamiltonian. In the absence of a scalar potential energy, Pauli Hamiltonian of a massive charged particle, i.e. an electron with the mass and charge,  is given in the form of \cite{Ajaib16}
\begin{eqnarray}
H_P=\frac{(\overrightarrow{\sigma}\cdot \overrightarrow{\Pi} )^{2}}{2m}=\frac{1}{2m}\left(\overrightarrow{\Pi}\cdot \overrightarrow{\Pi} +i\overrightarrow{\sigma} \cdot  (\overrightarrow{\Pi} \times \overrightarrow{\Pi})\right). \label{PauliHamiltonyen1}
\end{eqnarray}
Here, $\overrightarrow{\sigma}$ represents the
Pauli matrices, while $ \overrightarrow{\Pi}$ denotes the kinetic momentum terms which are defined by the minimal coupling of the canonical momentum, $\overrightarrow{p}$, and the vector potential, $\overrightarrow{A}=\overrightarrow{B}\times 
\overrightarrow{r}$, via $ \overrightarrow{\Pi} =\overrightarrow{p}-e \overrightarrow{A}$. Note that,  $\overrightarrow{B}$ is the external magnetic field.

\subsection{PO in commutative space}
We consider an oscillator potential energy, $\frac{1}{2}m\omega ^{2}r^{2}$, where $\omega$ is the oscillator frequency, and define the PO Hamiltonian in the form of
\begin{eqnarray}
H_{PO}&=&H_{P}+\frac{m\omega ^{2}}{2}r^{2}.
\end{eqnarray}
In a commutative space, we can examine the dynamic of an electron that is been the subject of the Pauli oscillator by the solving the following equation.
\begin{eqnarray}
\frac{1}{2m}\bigg[ \left(\overrightarrow{\Pi}\cdot \overrightarrow{\Pi} +i\overrightarrow{\sigma} \cdot  (\overrightarrow{\Pi} \times \overrightarrow{\Pi})
\right)+m^2\omega^{2}r^{2}\bigg] \Psi \left( \overrightarrow{r},t,s\right) =E\Psi \left( \overrightarrow{r},t,s\right). \label{POcom}
\end{eqnarray}
Alternatively, one can express Eq. \eqref{POcom} explicitly in the following form:
\begin{eqnarray}
&&\frac{1}{2m}\Bigg[ \left( \overrightarrow{p}-e \left(\overrightarrow{B}\times \overrightarrow{r} \right) \right) \cdot \left( \overrightarrow{p}-e \left(\overrightarrow{B}\times \overrightarrow{r} \right) \right) +i\overrightarrow{\sigma}\cdot \bigg( \left( \overrightarrow{p}-e \left(\overrightarrow{B}\times \overrightarrow{r} \right) \right) \times\left( \overrightarrow{p}-e \left(\overrightarrow{B}\times \overrightarrow{r} \right) \right) \bigg)  \nonumber \\ &+& m^{2} \omega^{2}r^{2}\Bigg]\Psi \left( \overrightarrow{r},t,s\right) = E\Psi \left( \overrightarrow{r},t,s\right).
\end{eqnarray}

\subsection{PO in a NC space}
In a NC spatial space, commuting operators are substituted with the noncommuting ones so that Eq. \eqref{POcom} is modified. Roughly speaking, the nonzero values of the new commutations of the coordinate operators cause additional terms to the commutative case which can be seen as a perturbative contribution \cite{Haouam20a}. To determine those terms, we have to state the NC space. In this manuscript, we consider the following spatial commutation relations of two NC operators \cite{Gamboa011}:
\begin{eqnarray}
\left[ \widehat{x}^{i},\widehat{y}^{j}\right] =i\theta ^{ij}.
\end{eqnarray}
Here, $\theta^{ij}$ is an antisymmetric tensor of spatial space with dimensions $\mathrm{(length)^2}$, which plays an analogous role to the reduced Planck constant of the usual quantum mechanics. In the NC space, unlike the commutative space, the product of any two infinitely small differentiable functions is expressed with the  Moyal-star product, that is  originally defined   by \cite{Moyal49}%
\begin{eqnarray}
\left( f\ast g\right) (x)=\exp \left[ \frac{i}{2}\theta _{\mu \nu }\partial
_{x_{\mu }}\partial _{y_{\nu }}\right] f(x)g(y)_{x=y}.
\end{eqnarray}
In the case of $\left[ \widehat{p}_{i},\widehat{p_{j}}\right] =0$, the NC quantum mechanics \ $H(p,x)\ast \Psi \left( \mathbf{x}\right)
=E\Psi \left( \mathbf{x}\right) $ reduces to usual one described by $H(\tilde{p},\tilde{x})\Psi \left( \mathbf{x}\right) =E\Psi \left( \mathbf{x}\right)$,  \cite{T6NC}. In other words, if one defines a new operators sets by  using the noncommuting coordinate operator in terms of following commuting coordinate operators, namely Bopp's shift \cite{Gouba16}, 
\begin{subequations}
\begin{eqnarray}
x_{i}\longrightarrow \widehat{x}_{i} &=&x_{i}-\frac{1}{2\hbar }\theta _{ij}p_{j}, \\
p_{i}\text{\ }\longrightarrow \widehat{p}_{i}&=&p_i,  
\end{eqnarray}
\end{subequations}
for $i=1,2$, then, the new operators obey the usual canonical commutation  relations  \cite{Gamboa02}. Here, $\theta ^{ij}$ is chosen as 
\begin{eqnarray}
\theta _{ij}=\epsilon _{ijk}\theta _{k}, \text{ and \ } \theta _{3}=\theta,
\end{eqnarray}
where $\epsilon _{ijk}$ is the Levi-Civita tensor. According to this choice, we obtain the PO equation in the commutative space as follows:
%
%
%
%
\begin{eqnarray}
&&\frac{1}{2m}\Bigg[ \left( \overrightarrow{p}-e \overrightarrow{B}%
\times \left( \overrightarrow{r}+\frac{\overrightarrow{\theta }\times 
\overrightarrow{p}}{2\hbar }\ \right) \right) .\left( \overrightarrow{p}-%
e \overrightarrow{B}\times \left( \overrightarrow{r}+\frac{%
\overrightarrow{\theta }\times \overrightarrow{p}}{2\hbar }\ \right) \right)  \nonumber \\
&+& i\overrightarrow{\sigma} \cdot \left( \overrightarrow{p}-e \overrightarrow{B}\times
\left( \overrightarrow{r}+\frac{\overrightarrow{\theta }\times 
\overrightarrow{p}}{2\hbar }\ \right) \right) \times \left( \overrightarrow{p%
}- e \overrightarrow{B}\times \left( \overrightarrow{r}+\frac{%
\overrightarrow{\theta }\times \overrightarrow{p}}{2\hbar }\ \right) \right)
 \nonumber \\
&+& m^2\omega ^{2}\left( \overrightarrow{r}+\frac{\overrightarrow{%
\theta }\times \overrightarrow{p}}{2\hbar }\ \right)\cdot \left( \overrightarrow{r}+\frac{\overrightarrow{%
\theta }\times \overrightarrow{p}}{2\hbar }\ \right)\Bigg] \Psi \left( 
\overrightarrow{r},t,s\right) =i\hbar \frac{\partial \Psi \left( \overrightarrow{r}%
,t,s\right) }{\partial t}.
\end{eqnarray}
For simplicity, we assume the orientation of the magnetic field along the z-axis. We take the NC gauge  
\begin{eqnarray}
\overrightarrow{A}=\overrightarrow{B}\times \overrightarrow{r}=\frac{B}{2}(-\widehat{y}i+\widehat{x}j), \quad A_0=0.
\end{eqnarray}
After some straightforward calculations, PO equation reduces to the following form: 
\begin{eqnarray}
\left[\frac{p_{x}^{2}+p_{y}^{2}}{2M}+\frac{p_{z}^{2}}{2m} +\frac{M\widetilde{\omega }^{2}}{2}\left( x^{2}+y^{2}\right)+\frac{m\omega ^{2}}{2}z^{2} -\text{\ }\eta L_{Z}-\frac{eB\hbar }{2m}\left( 1+\frac{eB}{4\hbar }\theta \right) \sigma _{z}\right] \Psi \left( \overrightarrow{r},t,
\mathbf{s}\right) =i\hbar \frac{\partial \Psi\left( \overrightarrow{r},t,s\right) }{\partial t}, \label{11}
\end{eqnarray}
where
\begin{subequations} \label{Consa}
\begin{eqnarray} 
M&=&m\left[\left( 1+\frac{eB \theta}{4\hbar} \right) ^{2}+\left( \frac{m\omega\theta }{2\hbar }\right) ^{2}\right]^{-1},  \\
\widetilde{\omega}&=&\omega\sqrt{\frac{m}{M}\left[ 1+  \left(\frac{eB}{2m\omega}\right)^{2}\right]}, \\
\eta &=& \frac{1}{2}\left[ \frac{eB}{m}\left( 1+\frac{eB\theta}{4\hbar } \right) +%
\frac{m\omega ^{2}\theta }{\hbar }\right]. 
\end{eqnarray}
\end{subequations}
Before we explore a solution to Eq. \eqref{11}, we briefly discuss the following cases:
\begin{itemize}
    
\item In a  commutative space, $\theta=0$, without an external magnetic field, $B=0$, we find $M = m$, $\widetilde{\omega} = \omega$ and $\eta=0$. Thus, PO equation transforms to the form of 
\begin{eqnarray}
\left[\frac{p_{x}^{2}+p_{y}^{2}+p_{z}^{2}}{2m}+\frac{m {\omega }^{2}}{2}\left( x^{2}+y^{2}+z^2\right) \right] \Psi  =i\hbar \frac{\partial \Psi }{\partial t}. \label{POord}
\end{eqnarray}

\item In a  commutative space, $\theta=0$, with the presence of an external magnetic field,  we find $M = m$, $\widetilde{\omega} = \omega \sqrt{1+ \left(\frac{e B }{2m} \right)^2}$ and $\eta=\frac{e B }{2m}$. Thus, PO equation transforms to the form of 
\begin{eqnarray}
\left[\frac{p_{x}^{2}+p_{y}^{2}+p_{z}^{2}}{2m}+\frac{m {\omega }^{2}}{2}\left( x^{2}+y^{2}+z^2\right)+ \frac{m \omega^2}{2}\left( \frac{e B }{2m\omega}\right)^2\left( x^{2}+y^{2}\right)-\frac{e B }{2m} ( L_z+\hbar\sigma _{z} )  \right] \Psi  =i\hbar \frac{\partial \Psi }{\partial t}. \,\,\,\,\,\,\,\,\,\,\,\, \label{POord2}
\end{eqnarray}

\item In a  NC space,  without an external magnetic field, $B=0$,  we find $M=m \left[1+\left( \frac{m\omega\theta }{2\hbar }\right) ^{2}\right]^{-1}$, $\widetilde{\omega} = \omega \sqrt{1+\left(\frac{m\omega \theta}{2\hbar}\right)^2}$ and $\eta=\frac{m\omega ^{2}\theta }{2\hbar }$. Thus, PO equation transforms to the form of 
\begin{eqnarray}
&&\left[\frac{p_{x}^{2}+p_{y}^{2}+p_{z}^{2}}{2m}+\frac{m \omega^{2}}{2}\left( x^{2}+y^{2}+z^{2}\right)+\left( \frac{m\omega\theta }{2\hbar }\right) ^{2}\left(\frac{p_{x}^{2}+p_{y}^{2}}{2m}\right) - \frac{m\omega ^{2}\theta }{2\hbar } L_{Z}\right] \Psi  =i\hbar \frac{\partial \Psi  }{\partial t}. \label{POord3}
\end{eqnarray}

Note that, in this case, the terms related to the deformation parameter can be taken as perturbation terms as we mentioned at the very beginning of this subsection.
\end{itemize}

\subsection{Solution}
In order to derive stationary PO equation out of Eq. \eqref{11}, we take the wave function as 
\begin{eqnarray}
\Psi \left( \overrightarrow{r}, t, s\right) =\exp \left({-\frac{iE t}{\hbar} }\right)\psi _{s}\left( \overrightarrow{r}\right).
\end{eqnarray}
Thus, we arrive at the following eigenvalue equation
%
%
\begin{eqnarray}
\left[ \frac{p_{x}^{2}+p_{y}^{2}}{2M}+\frac{M\widetilde{%
\omega }^{2}}{2}\left( x^{2}+y^{2}\right) +\frac{p_{z}^{2}}{2m}+\frac{m\omega
^{2}}{2}z^{2} -\eta L_{Z}\right] \psi _{s}\left( \overrightarrow{r}\right) =%
\left[ E+\frac{eB\hbar }{2m}\left( 1+\frac{eB \theta }{4\hbar}\right)
\sigma _{z}\right] \psi _{s}\left( \overrightarrow{r}\right). \label{Eq15}
\end{eqnarray}
%
%
%
Then, we introduce the cylindrical coordinate, $\left(r,\varphi,z\right)$,
and Eq. \eqref{Eq15} becomes
\begin{eqnarray}
&&\Bigg[ -\frac{\hbar ^{2}}{2M}\left( \frac{1}{r}\frac{\partial }{\partial r}\left(r%
\frac{\partial }{\partial r}\right)+\frac{1}{r^{2}}\frac{\partial ^{2}}{\partial
\varphi \text{\ }^{2}}\right) +\frac{M\widetilde{\omega }^{2}}{2}r^{2}-\frac{\hbar ^{2}}{2m}\frac{\partial ^{2}}{\partial z^{2}}
+\frac{m\omega ^{2}}{2}z^{2}
-\eta
\left( i\hbar \frac{\partial }{\partial \varphi \text{\ }}\right)\Bigg] \psi _{s}\left( \overrightarrow{r}\right) \nonumber \\
&=&\Bigg[ E+\frac{eB\hbar }{2m}%
\left( 1+\frac{eB \theta}{4\hbar } \right) \sigma _{z} \Bigg] \psi _{s}\left( 
\overrightarrow{r}\right). \label{Eq17}
\end{eqnarray}
Next, we express the two-component spinor to the following form
\begin{eqnarray}
\psi _{s}\left( \overrightarrow{r}\right) =\psi _{s}\left( r,\phi \right) =\chi
_{s}\exp (im_{l}\varphi )\frac{R(r)}{\sqrt{r}} Z(z), \quad s=+1,\,-1,
\label{Eq18}
\end{eqnarray}
where $\chi_{+1}^{T}=(1,0)$, and $\chi _{-1}^{T}=(0,1)$. By substituting Eq. \eqref{Eq18} into Eq. \eqref{Eq17}, we obtain two separate equations, namely the azimuthal and radial equations, respectively: 
\begin{subequations}
\begin{eqnarray}
\left[\frac{\hbar^2}{2m}\frac{\partial^2}{\partial z^2}-\frac{m\omega^2}{2}z^2+E_z\right]Z(z)&=&0 \label{EqAz} \\
\left[  \frac{d^{2}}{dr^{2}}+\frac{1}{r}\frac{d}{%
dr}-\frac{m_{l}^{2}}{r^{2}}-\frac{M^{2}\widetilde{\omega }^{2}}{\hbar ^{2}}%
r^{2}+\overline{E} \right] \frac{R(r)}{\sqrt{r}} &=&0,\label{Eqr}
\end{eqnarray}
\end{subequations}
%
where
\begin{eqnarray} 
\overline{E}&=&\frac{2M}{\hbar ^{2}}\left[ E+\frac{eB\hbar }{2m}\left( 1+\frac{eB\theta}{4\hbar 
} \right) s +\eta m_{l}\hbar -E_z
\right]. \label{Eustu}
\end{eqnarray}
The azimuthal one is the Schr\"odinger equation of the one dimensional harmonic oscillator, so that we have
\begin{eqnarray}
E_z=\hbar \omega\Big(n_{z}+\frac{1}{2}\Big)\,\,\,\,\, n_{z}=0,1,2,\cdots .
\end{eqnarray}
Note that the quantum number, $n_{z}$, is associated with the wave eigenfunction of the
harmonic oscillator along the external magnetic field. The corresponding eigenfunction of the azimuthal part reads
\begin{eqnarray}
Z(z)=C_{0}\exp \left( -\frac{m\omega}{2 \hbar} z^{2}\right) H_{n_{z}}\left( 
\sqrt{\frac{m\omega}{\hbar}}z\right),
\end{eqnarray}
where $C_{0}$ denotes the normalization constant and $ H_{n_{z}}\left( 
\sqrt{\frac{m\omega}{\hbar}}z\right)$ represents the Hermite polynomial. Then, we continue with the solution of the radial part. After a simple algebra, Eq. \eqref{Eqr} gives
\begin{eqnarray}
\left[ \frac{d^{2}}{dr^{2}}-\frac{M^{2}\widetilde{\omega }^{2}}{\hbar ^{2}}%
r^{2}-\frac{m_{l}^{2}-(1/4)}{r^{2}}+\overline{E}\right] R(r)=0.  \label{ara1}
\end{eqnarray}
Then, we consider the following solution
\bigskip 
\begin{eqnarray}
R(r)=e^{-\frac{\xi }{2}}\xi ^{k}W(\xi ), \label{eqr0}
\end{eqnarray}
where
\begin{subequations}
\begin{eqnarray}
\xi &=&\left( \frac{r}{a}\right) ^{2},  \\
a^{2}&=&\frac{%
\hbar }{M\widetilde{\omega }}=
\frac{\hbar }{m\omega} \sqrt{\frac{\left( 1+\frac{eB \theta}{4\hbar} \right) ^{2}+\left( \frac{m\omega\theta }{2\hbar }\right) ^{2}}{1+\left(\frac{eB}{2m\omega}\right)^2}}.
\label{akare}
\end{eqnarray}
\end{subequations}
Thus, Eq. \eqref{ara1} reduces to
\begin{eqnarray}
\left[ \xi \frac{d^{2}}{d\xi ^{2}}+\left( 2k+\frac{1}{2}-\xi \right) \frac{d%
}{d\xi }+n\right] W(\xi)=0, \label{Conf1}
\end{eqnarray}
where
\begin{subequations} \label{Ena}
\begin{eqnarray} 
k&=&\frac{1}{2}\left( \left\vert m_{l}\right\vert +\frac{1}{2}\right), \\
n&=&\frac{\varepsilon }{4}-\frac{1}{2}(\left\vert m_{l}\right\vert +1), \\
\varepsilon &=&\frac{\hbar }{M\widetilde{\omega}}\overline{E}.
\end{eqnarray}
\end{subequations}
We note that Eq. \eqref{Conf1} is the confluent hypergeometric equation whose solution is given in terms of two independent confluent hypergeometric functions. Among these solutions, we take the one that has a regular behavior
\begin{eqnarray}
W(\xi)= C_1 \,\, F(-n;\left\vert m_{l}\right\vert +1; \xi ), \label{eqr1}
\end{eqnarray}
where $C_1$ is the normalization constant. Then, we combine Eqs. \eqref{eqr0} and \eqref{eqr1} to build the radial solution 
\begin{eqnarray}
R(r)= C e^{-r^{2}/2a^{2}}r^{^{\left
\vert m_{l}\right\vert +1/2}}F\left(-n;\left\vert m_{l}\right\vert +1,\frac{r^{2}}{a^{2}}\right).
\end{eqnarray}
Here, $C$ denotes the normalization constant. Then, we obtain the wave function as follows:
\begin{eqnarray}
\psi _{n,m_{l},n_{z}}\left( \overrightarrow{r},t,s\right) =N \chi
_{s}\exp \left(im_{l}\varphi \right)\exp\left[- \left(-\frac{r^{2}}{2a^{2}}%
+\frac{m\omega}{2\hbar}z^2\right)\right] {\left(\frac{r^2}{a^2}\right)}^{\left\vert \frac{m_{l}}{2}\right\vert }F\left( -n;\left\vert
m{}_{l}\right\vert +1,\frac{r^{2}}{a^{2}}\right) H_{n_{z}}\left( \sqrt{\frac{m\omega}{\hbar}}z%
\right),\,\,
\end{eqnarray}
where $N$ is the normalization constant and its algebraic form is beyond our scope in this manuscript. We observe that the  contribution of the NC deformation appears in the parameter $a$ that is derived in Eq. \eqref{akare}. Before we investigate the quantization, we would like to present the value of this parameter in the three limit cases:

\begin{itemize}
    
    \item In a  commutative space, $\theta=0$, without an external magnetic field, $B=0$, 
\begin{eqnarray}
a^2= \frac{\hbar}{m\omega}.
\end{eqnarray}

\item In a  commutative space, $\theta=0$, with the presence of an external magnetic field,  
\begin{eqnarray}
a^2= \frac{\hbar}{m\omega}\left[1+ \left( \frac{eB}{2m\omega}\right)^{2}\right]^{-1/2}.
\end{eqnarray}

\item In a  NC space,  without an external magnetic field, $B=0$, 
\begin{eqnarray}
a^2&=&\frac{\hbar }{m\omega}\left[1+\left( \frac{m\omega\theta }{2\hbar }\right) ^{2}\right]^{1/2}.
\end{eqnarray}
\end{itemize}

%
For quantization, we use Eqs.  \eqref{Consa}, \eqref{Eustu}, and \eqref{Ena}.
We find
\begin{eqnarray}
E_{n,m_{l},n_{z}}&=&\hbar\omega\left[ \sqrt{\left[1+\left( \frac{eB}{2m \omega}\right)^{2}\right]\left[\left( 1+\frac{eB \theta}{4\hbar} \right) ^{2}+\left( \frac{m\omega\theta }{2\hbar }\right) ^{2}\right]}\left(
2n+\left\vert m_{l}\right\vert +1\right)+\Big(n_{z}+\frac{1}{2}\Big)\right] \nonumber \\ 
&-&\hbar\omega\left[ \frac{eB }{2m\omega}\left( 1+\frac{eB\theta }{4\hbar }\right)( s + m_{l}) +
\frac{m\omega\theta }{2\hbar }m_{l}\right]. \label{Enerji}
\end{eqnarray}

Before we proceed, let us examine the following cases:

\begin{itemize}
    
    \item In a  commutative space, $\theta=0$, without an external magnetic field, $B=0$, 
\begin{eqnarray}
E_{n,m_{l},n_{z}}=\hbar\omega\left[ \left( 2n+\left\vert m_{l}\right\vert +1\right) +\Big(n_{z}+\frac{1}{2}\Big) \right].
\end{eqnarray}
If one defines a principle quantum number as $\tilde{n}\equiv 2n+\left\vert m_{l}\right\vert+n_{z}$, then we arrive at
\begin{eqnarray}
E_{\tilde{n}}=\hbar\omega\left(\tilde{n} +\frac{3}{2}\right).
\end{eqnarray}

\item In a  commutative space, $\theta=0$, with the presence of an external magnetic field,  
\begin{eqnarray}
E_{n,m_{l},n_{z}}=\hbar\omega\left[ \sqrt{1+ \left( \frac{eB}{2m \omega}\right)^{2}}\left(
2n+\left\vert m_{l}\right\vert +1\right)+\Big(n_{z}+\frac{1}{2}\Big)-\frac{eB }{2m\omega}( s + m_{l}) \right].
\end{eqnarray}

\item In a  NC space,  without an external magnetic field, $B=0$, 
\begin{eqnarray}
E_{n,m_{l},n_{z}}=\hbar\omega\left[\sqrt{ 1+\left( \frac{m\omega\theta }{2\hbar }\right) ^{2}} \left(2n+\left\vert m_{l}\right\vert +1\right) + \Big(n_{z}+\frac{1}{2}\Big) +\Big(\frac{m\omega\theta }{2\hbar }\Big)m_{l} \right]. \label{eq41}
\end{eqnarray}

\end{itemize}

We observe that at critical magnetic field values, where the coefficient of $L_z$ vanishes, the effect of the NC space is able to counteract the effect of the normal Zeeman effect.  In order to determine the critical magnetic field values, we solve $\eta=0$. We find
\begin{eqnarray}
B_c &=& -\frac{2 \hbar}{e \theta} \left[1-\sqrt{1- \left(\frac{m \omega \theta}{\hbar}\right)^2}\right],
\end{eqnarray}
where we ignore the unphysical second root. At the critical value, the energy eigenvalue function reduces to
\begin{eqnarray}
E_{n,m_{l},n_{z}}&=&\hbar\omega\left[ \Big(\tilde{n}+\frac{3}{2}\Big)- \frac{m\omega\theta }{2\hbar }s \right].
\end{eqnarray}
Moreover for another critical value, 
\begin{eqnarray}
B_c &=& -\frac{4 \hbar}{e \theta},
\end{eqnarray}
the energy eigenfunction becomes
\begin{eqnarray}
E_{n,m_{l},n_{z}}&=&\hbar\omega\left[\sqrt{1+\left(\frac{m\omega \theta}{2\hbar}\right)^2} \left(
2n+\left\vert m_{l}\right\vert +1\right)+\Big(n_{z}+\frac{1}{2}\Big)+ \left(\frac{m\omega \theta}{2\hbar}\right)m_{l} \right].
\end{eqnarray}
This result is the same with Eq. \eqref{eq41}, even though the magnetic field value does not equal to zero. We conclude that the energy spectrum in the NC space becomes similar to the commutative space spectrum only in the critical values of magnetic field and NC space parameters.

\subsection{Upper bound value of the deformation}

In order to predict an upper bound value for the $\theta $ parameter, we use the $s-$ states of the energies from Eq. \ref{Enerji}. We Taylor expand up to the first order in deformation parameter, and we get
\begin{equation}
E_{n,0}=E_{n,0,0}^{\theta =0}+\hbar \omega \left[ \left( 2n+1\right) \frac{eB}{4\hbar }\sqrt{1+\left( \frac{eB}{2m\omega }\right) ^{2}}-\frac{\left(eB\right) ^{2}s}{8m\omega \hbar }\right]. \theta \label{MM1}
\end{equation}
with
\begin{equation}
E_{n,0,n_{z}}^{\theta =0}=\hbar \omega \left[ \left( 2n+1\right) \sqrt{1+\left( \frac{eB}{2m\omega }\right) ^{2}}+(n_{z}+\frac{1}{2})\right] -\frac{\hbar eBs}{2m}. \label{MM2}
\end{equation}
These two relations show that the deviation of the $n$-th energy level is caused by the modified commutation relations. This deviation can be expressed by:
\begin{equation}
\frac{\Delta E_{n,0}^{\theta }}{\hbar \omega }=\left[ \left( 2n+1\right) \frac{eB}{4\hbar }\sqrt{1+\left( \frac{eB}{2m\omega }\right) ^{2}}-\frac{\left( eB\right) ^{2}s}{8m\omega \hbar }\right] \theta \label{MM3}
\end{equation}
Next, we consider the experimental results of the electron cyclotron motion in a Penning trap. Accordingly, we take the cyclotron frequency of an  electron trapped in a magnetic field of strength $B$ as $\omega _{c}=eB/m_{e}$ (without deformation). Therefore,  we  write $m_{e}\hbar \omega _{c}=e\hbar B=10^{-52}kg^{2}m^{2}s^{-2}$ for a magnetic field of strength $B=6T$. We assume that only the deviations of the scale of $\hbar \omega _{c}$ can be detected at the level $n=10^{10}$, so that  $\Delta E_{n}<\hbar \omega _{c}$ (no perturbation of the $n$-th energy level is observed) \cite{Chang02}. We obtain the following constraint:
\begin{equation}
\theta <2.57\times 10^{-26}m^{2}. \label{MM4}
\end{equation}

\section{Thermal Quantities}\label{Thermal}

\begin{figure}
    \centering
    \subfigure[The partition function of the PO.]
    {
        \includegraphics[scale=1]{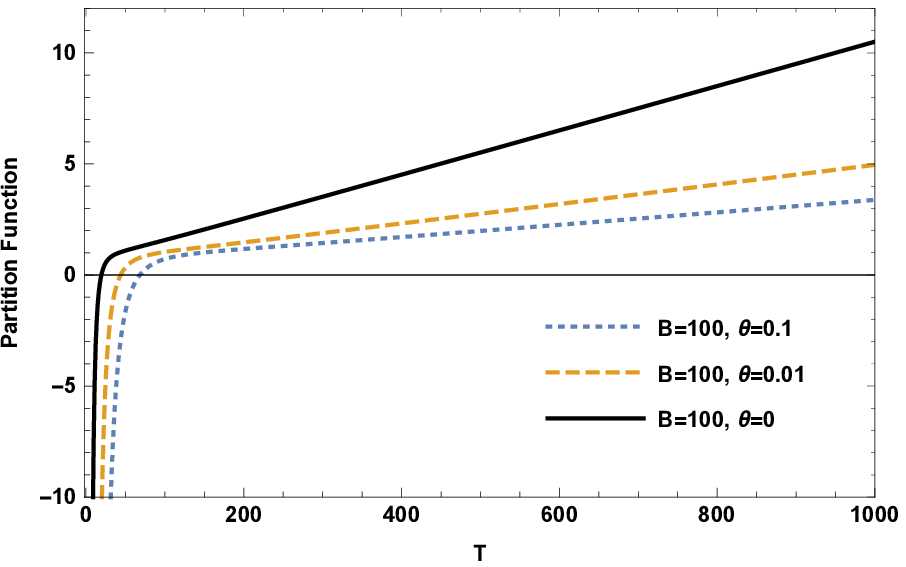}
        \label{fig:1a}
    }
    \\
    \subfigure[Effects of magnetic field in the NC space.]
    {
        \includegraphics[scale=0.9]{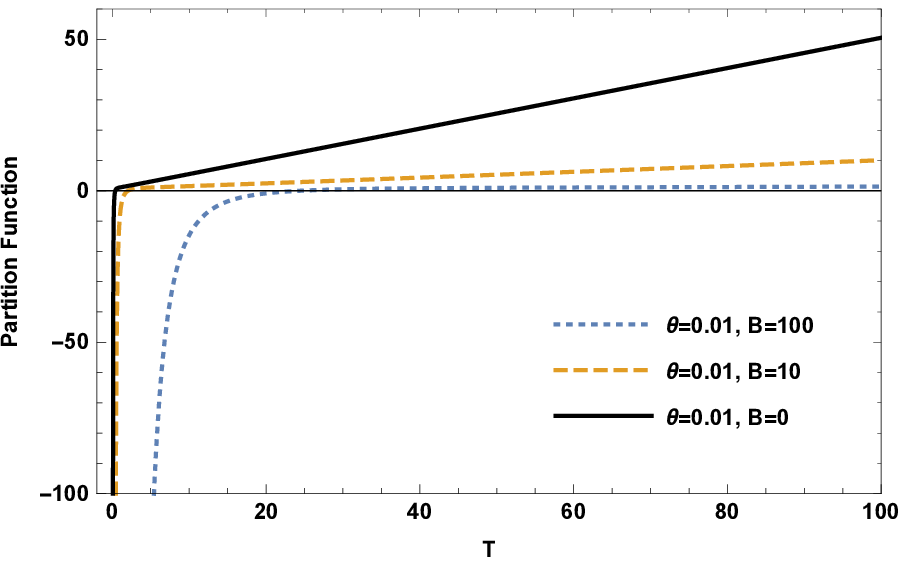}
        \label{fig:1b}
    }
    \subfigure[Effects of magnetic field in the commutative space.]
    {
        \includegraphics[scale=0.9]{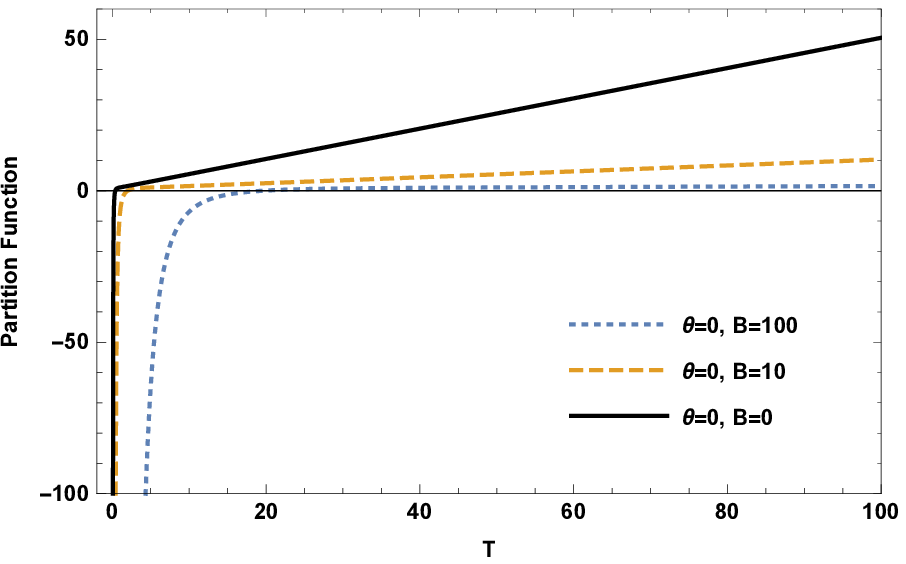}
        \label{fig:1c}
    }
    \caption{Partition function versus temperature for $\hbar=e=K_\beta=m=\omega=1$.}
    \label{fig:1}
\end{figure}

In this section, we examine the thermodynamic properties of the deformed PO at a finite temperature. For this, we consider the system to be in equilibrium with a reservoir at a constant temperature, $T$. Then, we use the well-known definition of the partition function:

\begin{eqnarray}
\mathcal{Z}(T)=\sum_{n=0}^{\infty }e^{\beta (E_{n}-E_{0})},
\end{eqnarray}
where $\beta =\frac{1}{K_{\beta }T}$, and $K_{\beta }$ is the Boltzmann constant. The obtained energy eigenfunction of the PO, given in Eq. \eqref{Enerji}, can be expressed as
\begin{eqnarray}
E_{n}=\alpha n+\lambda,
\end{eqnarray}
where
\begin{eqnarray}
\alpha =2\hbar\omega \sqrt{\left[1+\left( \frac{eB}{2m \omega}\right)^{2}\right]\left[\left( 1+\frac{eB \theta}{4\hbar} \right) ^{2}+\left( \frac{m\omega\theta }{2\hbar }\right) ^{2}\right]},
\end{eqnarray}
and
{\small
\begin{eqnarray}
\lambda &=&\hbar\omega\left[ \sqrt{\left[1+\left( \frac{eB}{2m \omega}\right)^{2}\right]\left[\left( 1+\frac{eB \theta}{4\hbar} \right) ^{2}+\left( \frac{m\omega\theta }{2\hbar }\right) ^{2}\right]}\left(
\left\vert m_{l}\right\vert +1\right)+\Big(n_{z}+\frac{1}{2}\Big)-\frac{eB }{2m\omega}\left( 1+\frac{eB\theta }{4\hbar }\right)( s + m_{l}) -
\frac{m\omega\theta }{2\hbar }m_{l}\right].\,\,\,\,\,\,\,\,
\end{eqnarray}}
Since $E_{0}$ is the fundamental state energy, namely $E_{0}=\lambda$, we immediately find $E_{n}-E_{0}=\alpha n$. Therefore, we express the partition function as
\begin{eqnarray}
\mathcal{Z}(T)=\sum_{n=0}^{\infty }e^{-\beta \alpha n}.
\end{eqnarray}
In order to calculate the partition function we employ the Euler-Maclaurin formula
\begin{eqnarray}
\sum_{n=0}^{\infty }f(n)=\frac{1}{2}f(0)+\int_{0}^{\infty
}f(x)dx-\sum_{p=1}^{\infty }\frac{1}{(2p)}B_{2p}f^{(2p-1)}(0),
\end{eqnarray}
where
\bigskip $B_{2p}$ indicates the Bernoulli numbers,  i.e.  $ B_{2}=\frac{1}{6}$, $B_{4}=-\frac{1}{30}$. Here,  $f^{(2p-1)}$ denotes the derivative of order $(2p-1)$.
In accordance with these facts, we get
\begin{subequations}
\begin{eqnarray}
\int_{0}^{\infty }f(x)dx &=&\int_{0}^{\infty }e^{-\beta (\alpha x)}dx=\frac{1}{\beta \alpha }, \\ 
f^{(1)}(0) &=& -\beta \alpha, \\
f^{(3)}(0) &=& -\beta ^{3}\alpha ^{3}.
\end{eqnarray}
\end{subequations}
Therefore, we obtain the  partition function in the following form
\begin{eqnarray}
\mathcal{Z}(\tau )\simeq\frac{1}{2}+\frac{\tau }{\alpha }+\frac{\alpha }{12\tau }-\frac{%
\alpha ^{3}}{120\tau ^{3}},
\end{eqnarray}
where $\tau =1/ \beta.$ Next, we use the partition to evaluate the thermal functions. At first, we obtain the free energy by recalling  $F=-\tau\ln (\mathcal{Z})$. 
\begin{eqnarray}
F=-\tau \ln \left(\frac{1}{2}+\frac{\tau }{\alpha }+\frac{\alpha }{12\tau }-%
\frac{\alpha ^{3}}{120\tau ^{3}}\right).
\end{eqnarray}
Then, we derive the average energy by employing $U=\tau ^{2}\frac{\partial \ln (\mathcal{Z})}{%
\partial \tau }$. We find
\begin{eqnarray}
U=\frac{120\tau ^{5}-10\alpha ^{2}\tau ^{3}+3\alpha ^{4}\tau }{60\alpha \tau
^{3}+120\tau ^{4}+10\alpha ^{2}\tau ^{2}-\alpha ^{4}}.
\end{eqnarray}
After that, we evaluate the reduced specific heat from $\frac{C}{K_B}=\frac{\partial U}{\partial \tau }$. We arrive at
\begin{eqnarray}
\frac{C}{K_B}=\frac{\left( 14400\tau ^{8}+14400\alpha \tau ^{7}+4800\alpha
^{2}\tau ^{6}-1780\alpha ^{4}\tau ^{4}-360\alpha ^{5}\tau ^{3}-3\alpha
^{8}\right) }{\left( 60\alpha \tau ^{3}+120\tau ^{4}+10\alpha ^{2}\tau
^{2}-\alpha ^{4}\right) ^{2}}.
\end{eqnarray}
Finally, we obtain the reduced entropy according to $\frac{S}{K_B}=-\frac{\partial F}{\partial \tau }$:
\begin{eqnarray}
\frac{S}{K_B}=\ln \left (\frac{1}{2}+\frac{\tau }{\alpha }+\frac{\alpha }{%
12\tau }-\frac{\alpha ^{3}}{120\tau ^{3}}\right)+\left( \frac{120\tau
^{4}-10\alpha ^{2}\tau ^{2}+3\alpha ^{4}}{60\alpha \tau ^{3}+120\tau
^{4}+10\alpha ^{2}\tau ^{2}-\alpha ^{4}}\right).
\end{eqnarray}
Using the following parameter values,  $\hbar=e=K_\beta=m=\omega=1$, we generate the plots of  derived statistical mechanical features versus temperature. In Fig. \ref{fig:1}, we present the partition function of the PO. We observe that in the higher valued NC parameter the partition function grows more slowly. We examine the effect of the magnetic field on the partition function in Figs. \ref{fig:1b} and \ref{fig:1c}. We see that in commutative and NC space the increase of magnetic field also changes the increase rate of the partition function. These modifications have similar characteristics. 

\begin{figure}
    \centering
    \subfigure[The Helmholtz free energy function of the PO.]
    {
        \includegraphics[scale=1]{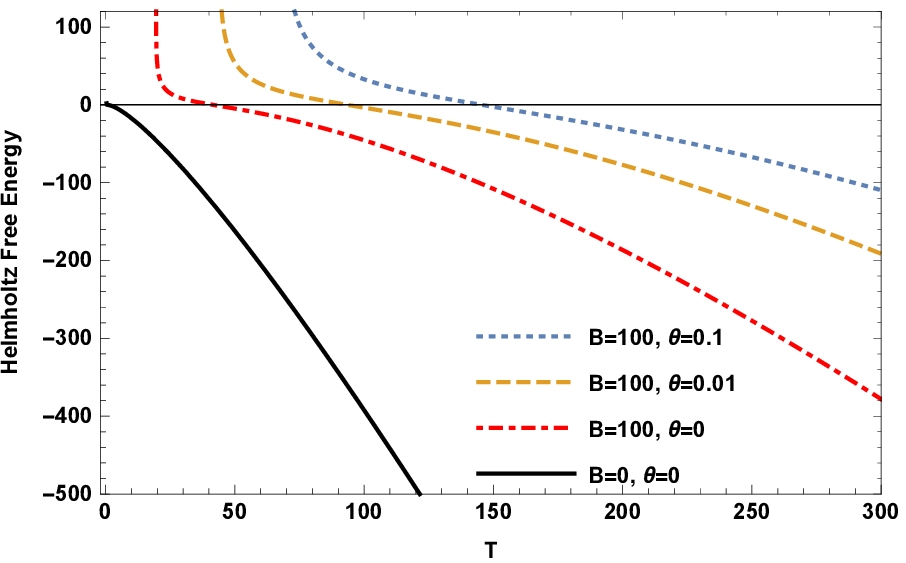}
        \label{fig:2a}
    }
    \\
    \subfigure[Effects of magnetic field in the NC space.]
    {
        \includegraphics[scale=0.9]{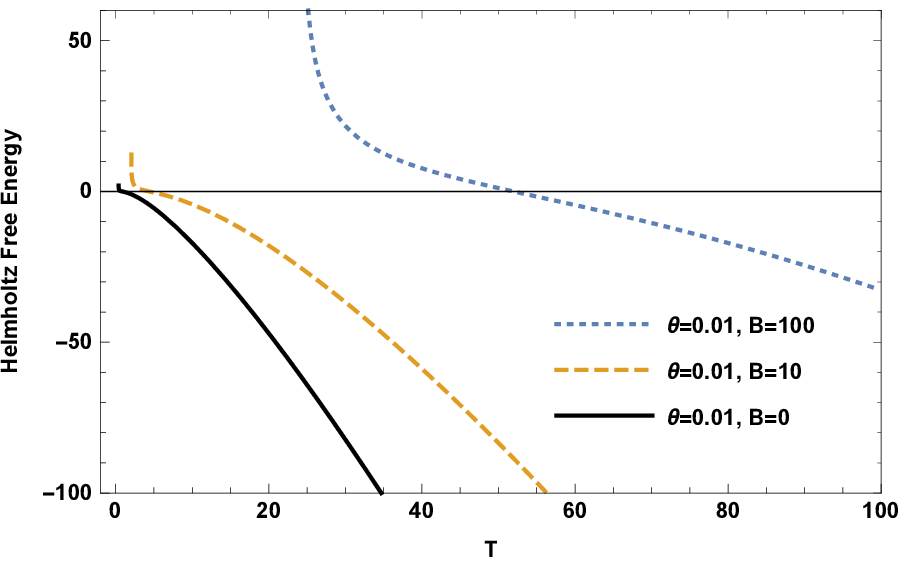}
        \label{fig:2b}
    }
    \subfigure[Effects of magnetic field in the commutative space.]
    {
        \includegraphics[scale=0.9]{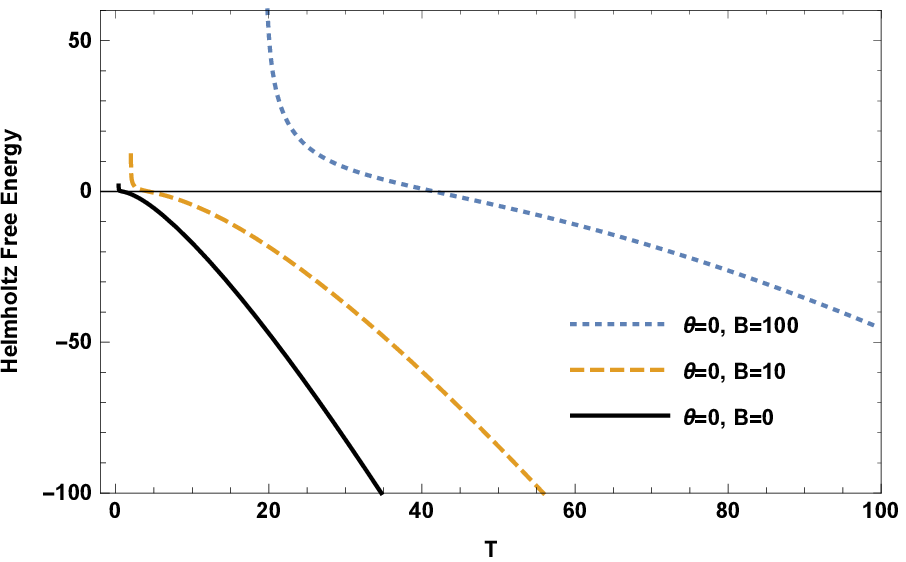}
        \label{fig:2c}
    }
    \caption{Helmholtz free energy function versus temperature for $\hbar=e=K_\beta=m=\omega=1$.}
    \label{fig:2}
\end{figure}

In Fig. \ref{fig:2} we illustrate the Helmholtz free energy function of the PO. In order to determine the difference, we plot the thermal function in the commutative space without an external magnetic field value in Fig. \ref{fig:2a}. We observe that the monotonic decrease of the function differs from the ones that are found in the NC space with the magnetic field.   This decrease becomes less in the larger NC parameter. Moreover, in this case, we do not observe a fall monotonically. The comparison of Figs. \ref{fig:2b} and \ref{fig:2c} points out that the effects of the magnetic field in the commutative and NC space are nearly the same. In both cases,  the Helmholtz free energy function has a higher value in the stronger magnetic field at a constant temperature value.

\begin{figure}
    \centering
    \subfigure[Internal energy function of the PO.]
    {
        \includegraphics[scale=1]{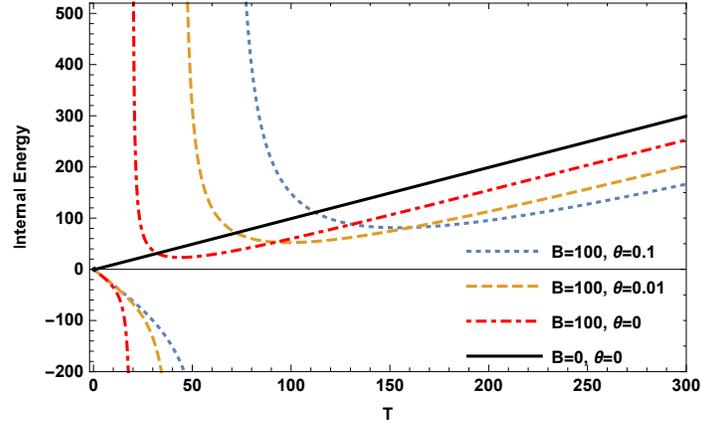}
        \label{fig:3a}
    }
    \\
    \subfigure[Effects of magnetic field in the NC space.]
    {
        \includegraphics[scale=0.9]{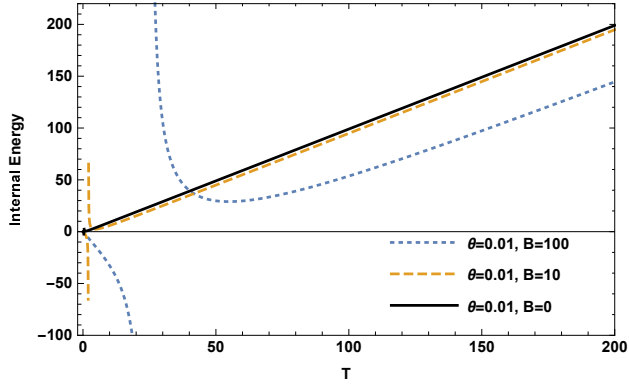}
        \label{fig:3b}
    }
    \subfigure[Effects of magnetic field in the commutative space.]
    {
        \includegraphics[scale=0.9]{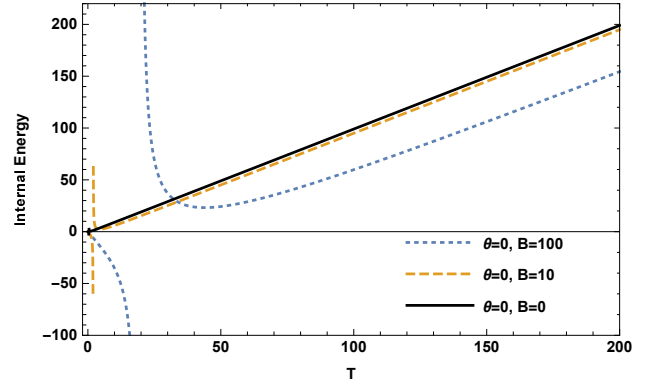}
        \label{fig:3c}
    }
    \caption{Internal energy function versus temperature for $\hbar=e=K_\beta=m=\omega=1$.}
    \label{fig:3}
\end{figure}

Next, we demonstrate the internal energy of the PO in Fig. \ref{fig:3}.
We notice a linear increase of the function as expected in the absence of a magnetic field in the commutative space. We observe a critical temperature value under the presence of an external magnetic field in commutative and NC space.  This critical temperature has a higher value for the bigger NC parameter. As seen in Fig. \ref{fig:3a}, the internal energy function decreases rapidly until the critical temperature. Furthermore, after the critical temperature value, we observe the same rapid decrease. However, after a turning point, the internal energy increases linearly. We detect a similar effect of the magnetic field on the internal energy of PO in Figs. \ref{fig:3b} and \ref{fig:3c}.

\begin{figure}
    \centering
    \subfigure[Specific heat function of the PO.]
    {
        \includegraphics[scale=1]{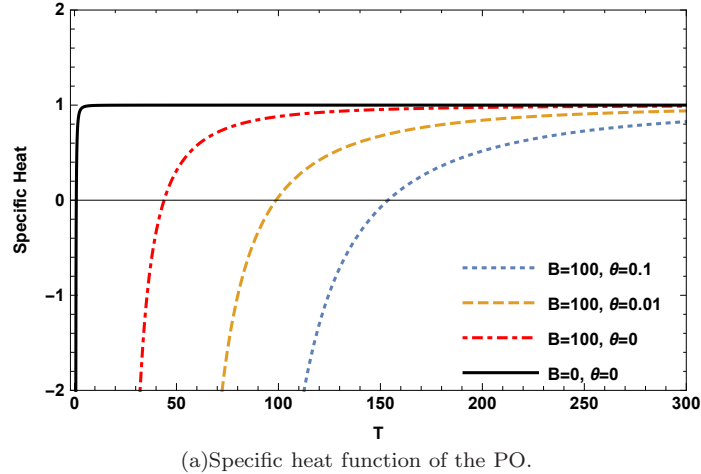}
        \label{fig:4a}
    }
    \\
    \subfigure[Effects of magnetic field in the NC space.]
    {
        \includegraphics[scale=0.9]{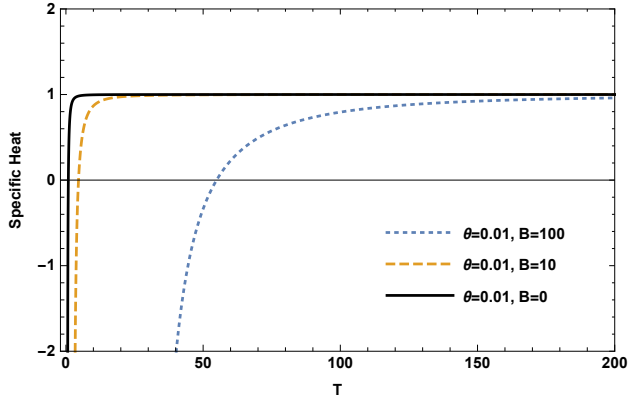}
        \label{fig:4b}
    }
    \subfigure[Effects of magnetic field in the commutative space.]
    {
        \includegraphics[scale=0.9]{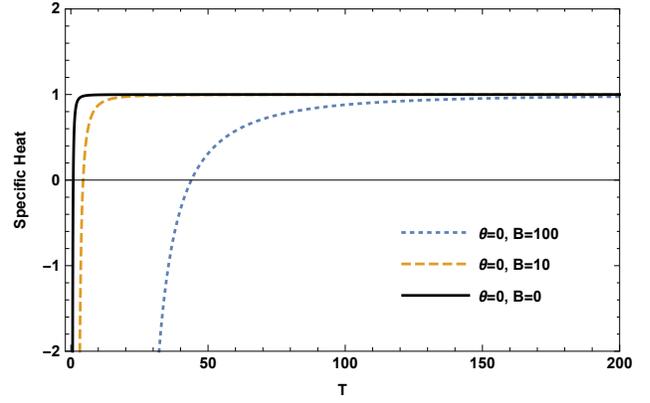}
        \label{fig:4c}
    }
    \caption{Specific heat function versus temperature for $\hbar=e=K_\beta=m=\omega=1$.}
    \label{fig:4}
\end{figure}

In Fig. \ref{fig:4}, we plot the specific heat of the PO. In all cases, we observe that this thermodynamic function tends to the same limit value. However, the external magnetic field and the NC space parameter determine the sharpness of the change. Fig \ref{fig:4a} shows that with the bigger NC space parameter the specific heat saturates at a higher temperature. Alike the previous thermal features, we observe a similar magnetic field effect in the commutative and NC commutative space. 

\begin{figure}
    \centering
    \subfigure[Entropy function of the PO.]
    {
        \includegraphics[scale=1]{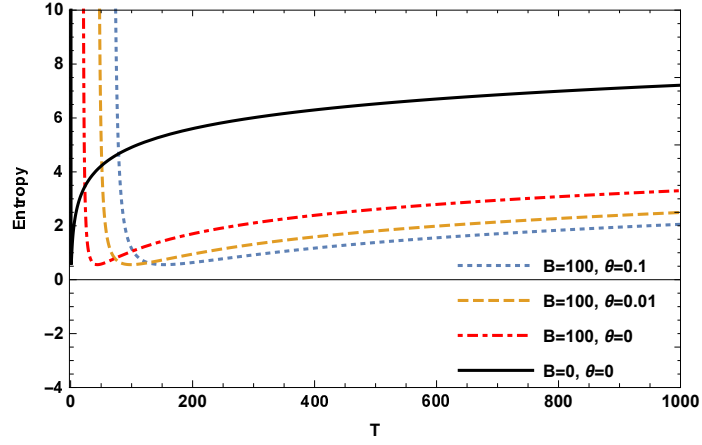}
        \label{fig:5a}
    }
    \\
    \subfigure[Effects of magnetic field in the NC space.]
    {
        \includegraphics[scale=0.9]{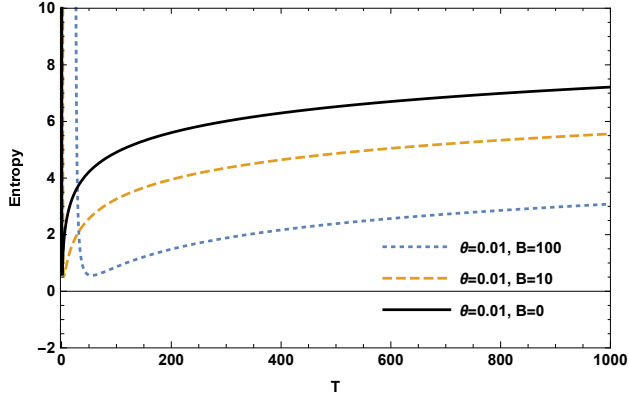}
        \label{fig:5b}
    }
    \subfigure[Effects of magnetic field in the commutative space.]
    {
        \includegraphics[scale=0.9]{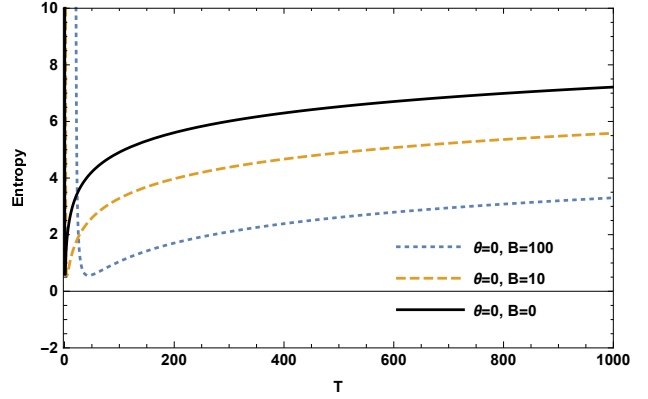}
        \label{fig:5c}
    }
    \caption{Entropy function versus temperature for $\hbar=e=K_\beta=m=\omega=1$.}
    \label{fig:5}
\end{figure}

Finally, we depict the entropy functions in Fig. \ref{fig:5}. We observe a rapid decrease in the relatively low-temperature values. After that, the entropy functions grow. However, this increase is less than the other increases of the greater NC parameter as seen in Fig. \ref{fig:5a}. We recognize the similar effects of the magnetic field on the entropy functions in both spaces.   

At this point, it is remarkable to note that these findings are tested by taking  $\theta \rightarrow 0$ the limit, where the ordinary expressions of the Pauli oscillator are obtained.

\section{Conclusion}\label{Concl}
In this work we have exposed an analytic study of the 2D Pauli oscillator in a noncommutative space and with the presence of a magnetic field. Written the Pauli equation of our system in the cylindrical coordinates $\left(r,\varphi,z\right)$, we found the explicit solutions of both the energy eigenvalues and the wave functions. The energy spectrum depends, as it should, on the deformation parameter $\theta$ which adds contributions coming from the interactions of the noncommutativity with both the angular momentum and the spin. These additional contributions are similar to those coming from the interaction of the system with the magnetic field and thus, we found that the effect of the noncommutative space is able to counteract the Zeeman effect when the magnetic field equal to a critical value $B_{c}=-\left(2\hbar / e\theta \right) \left[1\pm \sqrt{1-\left( m\omega\theta / \hbar\right)^{2}}\right] $. The same effect was found for the anomalous Zeeman effect and we got another critical value for the magnetic field $B'_{c}=-4c\hbar / e\theta$.

We have validated our spectrum by studying the limit $\theta \longrightarrow 0$ which give us the same results found for the ordinary space. We have also determined an experimental limit for the deformation parameter $\theta$ by using the results of the electron cyclotron motion in a Penning trap and thus, we have found that for a magnetic field of strength $B=6T$, we have the constraint $\theta <2.57\times 10^{-26}m^{2}$.

In the regime of high temperatures, we showed that the thermodynamic properties of our system have been also influenced by the noncommutative parameter $\theta$. The plots of these thermodynamic quantities illustrate that all the thermodynamic quantities decrease with increasing $\theta $ except the Helmholtz free energy.


\newpage

\newpage

\newpage

\newpage

\newpage

\end{document}